\documentclass[12pt]{article}
\usepackage{amsmath,amssymb,amsthm,enumerate}
\usepackage{graphicx}
\usepackage{epsfig}
\usepackage{bm}

\def\comm#1{}  
\newcommand{\nc}{\newcommand}  
\comm{
\unitlength 8mm
 \endcomm}

\def\H{{\mathbb H}} 

\def\hyt{\protect \mbox{${\mathbb H}_2\, $} }

\def\hyth{\mbox{$\H_3\, $}}
\def\HHN{\mbox{$\H_N$}}  
\def\HHt{{\hyt}}
\def\HH{\H}

\newcommand{\nn}{\nonumber}
\nc\vsk{\vskip 3mm}

\def\iw{{In\"on\"u-Wigner~}}
\def\iso{\simeq}

\nc\At{\widetilde{A}}
\nc\Ht{\widetilde{H}}
\nc\Qt{\widetilde{Q}}
\nc\Wt{\widetilde{W}}
\nc\Vt{\widetilde{V}}
\nc{\floor}[1]{\lfloor #1 \,\rfloor}
\nc{\wc}{\, , \,}   
\nc{\vn}{\mbox{v}}   
\newcommand{\crn}{\nn \\[2mm]}

\nc{\hrt}{{\heartsuit \, }}   

\nc{\Mb}{{\bf M}}

\def\Ac{{\cal A}}\def\Mc{{\cal M}}
\def\Nc{{\cal N}}

\def\cal{\mathcal}

\def\iw{{In\"on\"u-Wigner~}}

\def\g{{\mathfrak g}} 

\def\h{{\mathfrak h}} 

\def\w{{\mathfrak w}} 
\def\z{{\mathfrak z}}
\def\Lf{{\mathfrak L}}

\newcommand{\so}{{\mathfrak {so}}} 
\newcommand{\const}{\mbox{const.}}
\nc{\spf}{{\mathfrak {sp}}} 
\def\e{{\mathfrak e}} 

\nc{\pp}[1]{\mbox{~\H{\mbox{$\!\!\! #1$}}}}

\nc\ud{\underbrace}
\nc\bw{\bar{w}}
\nc\bu{\bar{u}}
\nc\by{\bar{y}}
\nc\Nb{{\bf N}}

\nc\ar{\acute{r}}
\nc\ax{\acute{x}}
\nc\au{\acute{u}}
\nc\aw{\acute{w}}
\nc\aaw{\acute{\!\acute{w}}}
\nc\aau{\mbox{\H{$ \!\!\! u$}}}

\nc\aar{\mbox{\H{$ \!\!\! r$}}}
\nc\aub{\acute{\ub}}

\nc{\nin}{\noindent}
\nc{\Sb}{{\bf S}}
\nc{\Jb}{{\bf J}}

\nc{\eq}{\!\!\! &=& \!\!\!}
\nc{\ptl}{\tilde{p}}  
\nc{\ut}{\tilde{u}}
\nc{\xt}{\tilde{x}}
\nc{\yt}{\tilde{y}}
\nc{\ttl}{\tilde{t}}

\def\bn{{\bf n}}

\newcommand{\ba}{\begin{array}}       
\newcommand{\ea}{\end{array}}

\nc{\leb}[1]{\left\{\ba{#1}}
\nc{\rip}{\ea \right.}

\newcommand{\ra}{\rightarrow}

\newcommand{\raz}{\rightarrow 0}

\newcommand{\Ra}{\Rightarrow }

\def\ga{\gamma}

\newcommand{\al}{\alpha}

\newcommand{\Sc}{{\cal S}}   

\def\half{\frac 1 2}

\def\eps{\epsilon}
\def\veps{\varepsilon}
\def\const{\mbox{const.}}
\def\al{\alpha}

\nc{\ph}{\varphi} 
\nc{\fii}{\phi}  

\def\lgs{\ifmmode{{\L(\g)^\s}}\else{\mbox{${\L(\g)^\s}$\ }}\fi}
\def\lgm{\ifmmode{{\L(\g)^\mu}}\else{\mbox{${\L(\g)^\mu}$\ }}\fi}
\def\lg{\ifmmode{{\L(\g)}}\else{\mbox{${\L(\g)}$\ }}\fi}
\def\ai{\ifmmode{{\alpha_i}}\else{\mbox{${\alpha_i}$\ }}\fi}  

\def\edd{\end{document}}
\def\hrt{\mbox{$\heartsuit$} }  
\nc\hn{ \mbox{${\bf h}^{\circ}$} }

\newcommand{\dff}{\noindent {\bf Definition.}\, }

\def\s{\ifmmode{{\sigma}}\else{\mbox{$\sigma$\ }}\fi}  
\def\z{\ifmmode{{\zeta}}\else{\mbox{$\zeta$\ }}\fi}  
\def\HN{\ifmmode{\HHN}\else{\HHN\ }\fi}  

\nc{\dg}{\dagger}

\def\sp#1{\langle #1 \rangle }

\newcommand{\rbbh}{\hat{\rbb}}

\newcommand{\xbbh}{\hat{\xbb}}
\newcommand{\ybbh}{\hat{\ybb}}

\newcommand{\ub}{{\bf u}}

\nc\bnabla{\mbox{\boldmath{$\nabla$}}}
\nc\bzero{\mbox{\boldmath{$0$}}}
\nc{\betab}{\mbox{\boldmath{$\beta$}}}
\nc{\betabh}{\hat{\mbox{\boldmath{$\beta$}}}}
\nc{\phb}{\mbox{\boldmath{$\phi$}}}
\nc{\psib}{\mbox{\boldmath{$\psi$}}}
\nc{\phib}{\mbox{\boldmath{$\phi$}}}
\nc{\phibh}{\mbox{\boldmath{$\hat{\phi}$}}}  
\nc{\phbh}{\hat{\mbox{\boldmath{$\varphi$}}}}

\nc{\rhob}{{\mbox{\boldmath{$\rho$}}}}
\nc{\rhobh}{\hat{\mbox{\boldmath{$\rho$}}}}
\nc{\sigmab}{\mbox{\boldmath{$\sigma$}}}
\nc{\thb}{\mbox{\boldmath{$\theta$}}}
\nc{\thbh}{\hat{\mbox{\boldmath{$\theta$}}}}

\def\HH{{\mbox{${H \! \!  I}$}}}
\def\HA{\ifmmode{\HH}\else{\HH\ }\fi}  
\def\and{\quad \mbox{ and } \quad }

\nc\crr{commutation relation}
\nc{\alb}{\mbox{\boldmath{$\alpha$}}}  
\nc{\xib}{\mbox{\boldmath{$\xi$}}}  
\nc{\pib}{\mbox{\boldmath{$\pi$}}}  
\nc{\dpib}{\dot{\pib}}  
\nc{\Pib}{\mbox{\boldmath{$\Pi$}}}  
\def\Bb{\mbox{\boldmath{$B$}}}

\nc\Ab{\mbox{\boldmath{$A$}}}  
\nc\Abt{\tilde{\Ab}}
\nc\Eb{\mbox{\boldmath{$E$}}}  
\nc\Db{\mbox{\boldmath{$D$}}}  
\def\Fbb{\mbox{\boldmath{$F$}}}
\def\Jb{\mbox{\boldmath{$J$}}}
\def\Kb{\mbox{\boldmath{$K$}}}
\def\Lbb{\mbox{\boldmath{$L$}}}
\def\ab{\mbox{\boldmath{$a$}}}
\def\db{\mbox{\boldmath{$d$}}}
\def\eb{\mbox{\boldmath{$e$}}}
\def\bb{\mbox{\boldmath{$b$}}}\def\cb{\mbox{\boldmath{$c$}}}

\nc{\intt}{{\int \! \int}}

\nc{\rb}{{\bf r}}
\nc{\qb}{{\bf q}}
\nc{\xb}{{\bf x}}
\nc{\vb}{{\bf v}}
\nc{\Fb}{{\bf F}}
\nc{\Ub}{{\bf U}}

\def\pbb{\mbox{\boldmath{$p$}}}
\def\rbb{\mbox{\boldmath{$r$}}}

\def\xbb{\mbox{\boldmath{$x$}}}\def\ybb{\mbox{\boldmath{$y$}}}

\def\x{\times}

\def\AA{\ifmmode{\mbox{\boldmath{$A$}}}\else{\A\ }\fi}  
\def\A{\ifmmode{\Ab}\else{\Ab\ }\fi}  
\def\B{\ifmmode{\Bb}\else{\Bb\ }\fi}  
\def\F{\ifmmode{\Fbb}\else{\Fbb\ }\fi}  
\nc{\Lb}{\ifmmode{\Lbb}\else{\Lbb\ }\fi}  
\def\L{{\cal L}}   
\def\K{\ifmmode{\Kb}\else{\Kb\ }\fi}  
\def\M{\ifmmode{\Mb}\else{\Mb\ }\fi}  %

\def\gbhh{{\mbox{\boldmath{$\widehat{g}$}}}}
\def\hbhh{{\widehat{\mbox{\boldmath{$h$}}}}}
\def\nbhh{{\widehat{\mbox{\boldmath{$n$}}}}}

\def\hbh{\ifmmode{\hbhh}\else{\hbhh\ }\fi}  %
\def\gbh{\ifmmode{\gbhh}\else{\gbhh\ }\fi}  %
\def\nbh{\ifmmode{\nbhh}\else{\nbhh\ }\fi}  %

\def\sgn {\mbox{ sgn }}

\newcommand{\lb}[1]{\label{#1}}
\newcommand{\rf}[1]{(\ref{#1})}
\newcommand{\be}{\begin{equation}}

\newcommand{\ee}{\end{equation}}
\newcommand{\el}[1]{\label{#1}\end{equation}}
\newcommand{\erl}[1]{\label{#1}\end{eqnarray}}
\newcommand{\egl}[1]{\label{#1}\end{gather}}
\newcommand{\bg}{\begin{gather}}
\newcommand{\egat}{\end{gather}}
\newcommand{\br}{\begin{eqnarray}}
\newcommand{\er}{\end{eqnarray}}

\newcommand{\for}{\qquad \mbox{ for}\quad}
\newcommand{\aand}{\quad \mbox{ and}\quad}

\newcommand{\where}{\quad \mbox{ where}\quad}
\newcommand{\with}{\quad \mbox{ with}\quad}

\newcommand{\ie}{{\em i.e.~}}
\newcommand{\sbs}{\subsection}
\newcommand{\etal}{{\em et.\ al.~}}

\begin{document}

\hfill {\bf to be published in JMP in August, 2006} \\[2mm]

\begin{center}

{\LARGE \bf Contraction of broken symmetries via Kac-Moody formalism}\\[5mm]

{\bf Jamil Daboul}\footnote{On Sabbatical
leave from the Physics Department, Ben Gurion University of the Negev,
84105 Beer Sheva, Israel (e-mail: jdaboul@gmail.com)} \\[15pt]
Centro de Ciencias F\'{\i}sicas
\\ Universidad Nacional Aut\'onoma de
M\'exico\\  Apartado Postal 48-3, Cuernavaca, Morelos 62251\\[3pt]\today

\end{center}

\bigskip

\begin{abstract}

I investigate contractions via Kac-Moody formalism. In particular, I show how
the symmetry algebra of the standard 2-D Kepler system, which was identified by
Daboul and Slodowy as an infinite-dimensional Kac-Moody
loop algebra, and was denoted by ${\mathbb H}_2\, $, gets reduced by the
symmetry breaking term, defined by the Hamiltonian
\[
H(\beta)= \frac 1 {2m} (p_1^2+p_2^2)- \frac \alpha r
- \beta  \ r^{-1/2} \cos ((\varphi-\gamma)/2)~.
\]
For this $H (\beta)$ I define two symmetry
loop algebras ${\mathfrak L}_{i}(\beta),~i=1,2$, by choosing the
`basic generators' differently. These ${\mathfrak L}_{i}(\beta)$ can be mapped
isomorphically onto subalgebras of ${\mathbb H}_2\, $, of codimension 2 or 3, revealing
the reduction of symmetry.
Both factor algebras ${\mathfrak L}_i(\beta)/I_i(E,\beta)$, relative to the
corresponding energy-dependent ideals $I_i(E,\beta)$, are isomorphic
to ${\mathfrak so}(3)$ and ${\mathfrak so}(2,1)$
for $E<0$ and $E>0$, respectively, just as for the pure Kepler case. However,
they yield two different non-standard contractions as $E \rightarrow 0$, namely
to the Heisenberg-Weyl algebra ${\mathfrak h}_3={\mathfrak w}_1$  or to an abelian Lie algebra, instead
of the Euclidean algebra ${\mathfrak e}(2)$ for the pure Kepler case.
The above example suggests a general procedure for defining generalized
contractions, and also illustrates the {\em `deformation contraction
hysteresis'}, where contraction which involve two contraction parameters can yield
different contracted algebras, if the limits are carried out in different order.

\end{abstract}


\newpage

\section{Introduction}

In 1926 Pauli \cite{pauli} obtained the energy levels of the relativistic
hydrogen atom algebraically, by using the conserved angular-momentum
$\Lb=\rbb \x \pbb$
and the Hermitian form of the Laplace-Runge-Lenz vector
\be
\Ab =\half [\pbb\x \Lb - \Lb\x \pbb] - m\al \rbbh
\el{rl}
The commutation relations among their components are given by
\br
[L_i, L_j] &=& i \hbar\ \eps_{ijk} L_k ~,  \qquad i,j,k=1,2,3  \cr
[L_i, A_j] &=& i \hbar\ \eps_{ijk} A_k ~,   \cr
[A_i, A_j] &=& -i 2mH \ \hbar\ \eps_{ijk} L_k ~,
\erl{cr}
where $H$ is the Hamiltonian of the non-relativistic 3D hydrogen atom. The
commutation relations in \rf{cr} do not define
a closed algebra, since the $H$ on the rhs of \rf{cr} is
an {\em operator} and not a number.
To obtain nevertheless closed algebras physicists for
seventy years have replaced the Hamiltonian $H$ by its eigenvalues $E$,
and thus obtained three different identifications of the symmetry algebra of the hydrogen
atom, namely $\so(4), \so(3,1)$ and $\e(3)$, for $E<0, E>0$ and $E=0$,
respectively \cite{sud}. The same conclusion can be reached by formally
`normalizing' the Runge-Lenz vector $\Ab$ by dividing it by $\sqrt{2m|H|}$,
but the resulting quotient vector becomes infinite for $H=0$.

Instead of the above `conventional procedure' , Daboul and Slodowy \cite{dsd}
showed that one can obtain a single
closed algebra based on the commutation relations \rf{cr}. This algebra
is spanned  by the following infinite set of generators
\be
\hyth := \{ h^n L_i, h^n A_i~|~ i=1,2,3, ~n=0,1,\ldots\}~, \where h:=- 2mH~.
\el{lz}
The algebra $\hyth$ and its generalizations $\HN$, the symmetry algebras
of the N-dimensional hydrogen atom, were
identified \cite{dsd,dd} as positive loop algebras of twisted or untwisted
Kac-Moody algebras  \cite{kac,fs}, for $N$ odd or even, respectively.
They were called the {\em hydrogen algebras}.
The above formalism will be reviewed in Sec. 2, and applied to
\hyt, the hydrogen algebra of the standard 2D Kepler system, defined by
the Hamiltonian $H_0$ of Eq. \rf{h3} below.

The algebras $\HN$ depend on the Hamiltonian $H$, but {\em not} on its energy
eigenvalues $E$. However, one can reproduce the usual three corresponding
finite-dimensional algebras, $\so(N+1), \so(N,1)$ and $\e(N)$, as factor algebras
$\HN/I(E)$ relative to energy-dependent ideals $I(E)$; The ideals and
factor-algebra formalism will be discussed and applied to \hyt in Sec. 2.1.

In the present paper I investigate {\em what happens to the algebra \hyt
and its factor algebra, if the original symmetry of the 2D hydrogen atom is
broken}. In particular,
I shall study the following Hamiltonian
\be
H := H_0 - \beta r^{-1/2} \cos \left[\half(\ph-\ga)\right]~, \qquad
\mbox{($\ga=0$ in the present paper)}
\el{h4}
where $H_0$ is the Hamiltonian of the 2-dimensional Kepler problem
\be
H_0 := \frac 1 {2m} (p_1^2+p_2^2)
- \frac \al r =
\frac  1 {2m} \left(p_r^2+ \frac {p_\ph^2}{r^2}\right) - \frac \al r~.
\el{h3}
Throughout this paper I shall set the phase angle $\ga$ in \rf{h4} equal to zero, since
it can always be removed by appropriate choice of the coordinate system (See however the
discussion in section 5 below).

The Hamiltonian  \rf{h4} has an interesting history: It was discovered
by Winternitz {\em et.\ al.~}\cite{win} already in 1967
in their systematic search for super-integrable systems.
It was
also derived in a more general complex form
by T. Sen \cite [Eq. (3.14)] {sen} in 1987.\\
The symmetry of \rf{h4} was originally studied by Gorringe and Leach
\cite{gl} in 1993 and recently reviewed by
Leach and Flessas \cite[\S 3.3]{lf} (see also \cite{kmp}).
The above authors followed the conventional method
and found that the symmetry algebras of \rf{h4} are $\so(3)$ and $\so(2,1)$
for $E<0$ and $E>0$, exactly as for the pure 2D Kepler problem \rf{h3}.
However, for  $E=0$ they obtained the Heisenberg-Weyl algebra $\h_3=\w_1$
(which they denoted by $W(3,1)$)  \cite{gl,lf}, instead of the Euclidean algebra
$\e(2)$ for the Kepler case \rf{h3}.

This result was intriguing, since the symmetry
breaking does not affect the symmetry for $E\ne 0$, and only affect it for
$E=0$. And I wondered whether and how the above
type of symmetry breaking can be treated via the Kac-Moody formalism.
It turned out, that the symmetry algebra of \rf{h4} can be treated, via the
Kac-Moody formalism, similar to
the pure Kepler case, with some important modifications. For example, it is
possible to describe the symmetry algebra of \rf{h4} by two loop algebras,
$\Lf_1$ and $\Lf_2$, depending on the choice of the
{\em `basic generators'}.
It is remarkable that these two algebras can be mapped onto subalgebras of
\hyt of codimension 2 and 3, \ie \hyt is larger than these image
subalgebras by only 2 and 3 generators, out of infinitely many.
The `missing' generators are manifestations of the symmetry breaking.

Moreover, I will show that the factor algebras $\Lf_i/I_i(E,\beta)$ relative to
the corresponding energy-dependent ideals yield different types of
{\em contractions} \cite{iw,gil,cla}, which are included in table 1.
This result is important, since the contraction procedure for the above
specific system can be generalized to other algebras, as discussed in
the summary section.

In Sec. 2 I review In\"on\"u-Wigner contraction and its generalization and in
Sec. 3 I review the construction of the hydrogen algebra \hyt
for the pure 2-dimensional Kepler problem \rf{h3}.
In sections 4 and 5 I construct two loop algebras $\Lf_1$ and $\Lf_2$
for the system \rf{h4} and their factor algebras $\Lf_i/I_i(E,\beta)$.
In Sec. 6 I map the $\Lf_i$ onto subalgebras \hyt, and as I already noted,
I shall suggest a general procedure for defining contraction via
Kac-Moody formalism and then give some conclusions.

\section{Review of  generalized In\"on\"u-Wigner contraction}
\label{reviewc}

There are many formulations of contractions \cite{iw, gil}. I shall give my
own definition and notation:

\dff Let $\g:=\sp{X_a,C_{ab}^c }$ be a finite-dimensional Lie algebra with a
basis $X_a, a=1,2,\ldots, N$ and structure constants $C_{ab}^c $, and let the
parameter-dependent Lie algebra $\g^\eps:=\sp{X_a^\eps ,C_{ab}^c(\eps) }$ be defined,
such that the one-to-one linear map $f_\eps$ between $\g$ and $\g^\eps$,
\be
f_\eps:  \g \mapsto \g^\eps, \qquad f_\eps(X_a)=  \eps^{- n_a} \, X_a^\eps
\el{map}
is an isomorphism of Lie algebras as long as $\eps \ne 0$. If the powers $n_a$ satisfy the
condition,
\be
{n_a+n_b\ge n_c}
\el{cnd}
then the limit algebra $\g^0=\sp{X_a^0,C_{ab}^c(0) }$ with the structure constants
\[
C_{ab}^c (0): =\lim_{\eps\raz} C_{ab}^c (\eps)
\]
exists and it is called the {\em contracted algebra}.   I shall refer to $\g^\eps$ as
the {\em contracting algebra} and to its generators $X_a^\eps$ as the {\em contracting
generators}.

It is important to emphasize that {\em $X_a^\eps$ and $X_a^0$ denote the generators of
the Lie algebras $\g^\eps$ and $\g^0$
which are defined via the structure constants $C_{ab}^c (\eps)$ and $C_{ab}^c (0)$,
respectively}.
Therefore, the {\em $X_a^0$ are NOT to be regarded as the limits of $X_a^\eps$ for
$\eps\raz$.}
Thus,  the $X_a^0$ {\em always} exist, by definition, as generators of the contracted
algebra $\g^0$, even though representations
$r(X_a^\eps)$ of $\g^\eps$ might exist with  some of the generators having vanishing
limits, \ie $\lim_{\eps \raz} r(X_b^\eps)=0$. Such representations could be called
{\em not saved or un-saved}. Otherwise, they are called {\em saved representations}
\cite{sal}. Actually in section 5 I shall give a realization of a saved representation of
an algebra whose
contraction yields an abelian algebra, \ie $C_{ab}^c=0$ for all $a, b, c$. For this
contraction even the adjoint representation \cite {gil} is not saved.

Usually $X_a$ is used also to denote  the contracting and contracted generators $X_a^\eps$
and $X_a^0$ \cite{gil}. This convention is probably used to avoid confusing $X_a^0$ as the
limit of  $X_a^\eps$ for $\eps \raz$. To distinguish the algebras $\g^\eps$ from
$\g$ and $\g^0$
one attaches an index $\eps$ to the commutators $[,]_\eps$, as it is done in \rf{cccc} below.
I find this usual notation confusing, since $X_a^0$ are the generators of a different
algebra $\g^0$. I prefer
attaching the $\eps$ to the generators but keep the commutator
symbol $[,]$ unchanged. This notation
is more useful and user-friendly, especially if matrix representations exist, since one uses
$[A,B]=AB-BA$ and the standard matrix multiplication, whether the matrices $A$ and $B$
represent generators of the original or the contracted algebras.

In contrast to the formal definition of $X_a^0$, the limits
$r(X_b^0):=\lim_{\eps \raz} r(X_b^\eps)$
of {\em representations or realizations}  $r(X_a^\eps)$, if they exist, should
satisfy the commutation relations of $\g^0$, although some of these
representations may not be saved.

The condition \rf{cnd} is necessary and sufficient to make the limit algebra $\g^0$
well defined. It insures that the {\em contracting structure constants}
$C_{ab}^c (\eps) $, defined by
\br
\sum_{c=1}^N  \, C_{ab}^c (\eps)\, X_c^\eps &:=& [X_a^\eps, X_b^\eps]_\eps =
\eps^{n_a+n_b}[f_\eps(X_a), f_\eps(X_b)]_\eps =\eps^{n_a+n_b} f_\eps([X_a, X_b])\cr
&=& \eps^{n_a+n_b} f_\eps \left(\sum_{c=1}^N C_{ab}^c  X_c\right)
= \sum_{c=1}^N \eps^{n_a+n_b-n_c} \, C_{ab}^c X_c^\eps ,
\erl{cccc}
have finite limits for $\eps\raz$.

\comm{
\dff {Let $\g:=\sp{X_a,C_{ab}^c }$ be a finite-dimensional Lie algebra with a
basis $X_a, a=1,2,\ldots, N$ and structure constants $C_{ab}^c $, and let the
parameter-dependent Lie algebra $\g^\eps:=\sp{X_a,C_{ab}^c(\eps) }$ be defined, such that
the one-to-one linear map $f_\eps$ between $\g$ and $\g^\eps$,
\be
f_\eps:  \g \mapsto \g^\eps, \qquad f_\eps(X_a)=  \eps^{- n_a} \, X_a
\el{map}
is an isomorphism of Lie algebras as long as $\eps \ne 0$. If the powers $n_a$ satisfy
the condition,
\be
{n_a+n_b\ge n_c}
\el{cnd}
then the limit algebra $\g^0=\sp{X_a,C_{ab}^c(0) }$ with the structure constants
$C_{ab}^c (0): =\lim_{\eps\raz} C_{ab}^c (\eps)$ exists and
it is called the {\em contracted algebra}.   I shall refer to $\g^\eps$ as the
 {\em contracting algebra}.}
The condition \rf{cnd} is necessary and sufficient to make the limit algebra $\g^0$
well defined. It is equivalent to that the {\em contracting structure constants}
$C_{ab}^c (\eps) $, defined by
\br
\sum_{c=1}^N  \, C_{ab}^c (\eps)\, X_c:=[X_a, X_b]^\eps &=&   \eps^{n_a+n_b}
[f_\eps(X_a), f_\eps(X_b)]^\eps =\eps^{n_a+n_b} f_\eps([X_a, X_b])\cr
&=& \sum_{c=1}^N \eps^{n_a+n_b-n_c} \, C_{ab}^c X_c ,
\erl{cccc}
have finite limits for $\eps\raz$.
\endcomm}

The {\em In\"on\"u-Wigner contraction} is a special case of the above definition, where
\br
n_i&=& 0 \for i=1,2, \ldots ,M, \aand \cr
n_\al &=& \const >0  \for \al=M+1,M+2, \ldots N~.
\erl{iwn}
In this case, and by choosing $\const=1$ for convenience, we obtain for $\eps \raz$:
\br
[X_i^\eps, X_j^\eps] &=&  \sum_{k=1}^M C_{ij}^k X_k^\eps \cr
&\Ra&  \sum_{k=1}^M C_{ij}^k X_k^0~, \where X_k^0 := \lim_{\eps\raz} X_k^\eps  \lb{VR}\\
~[X_i^\eps, X_\al^\eps ] &=&  \sum_{k=1}^M \eps~ C_{i\al}^k X_k^\eps +
\sum_{\beta=M+1}^N C_{i\al}^\beta X_\beta^\eps \cr
&\Ra& \sum_{\beta=M+1}^N C_{i\al}^\beta X_\beta^0 \lb{CR} \\
~[X_\al^\eps, X_\beta^\eps] &=&  \sum_{k=1}^M \eps^2~ C_{\al\beta}^k X_k^\eps +
\sum_{\ga=M+1}^N \eps~ C_{\al\beta}^\ga X_\ga^\eps \cr
&\Ra & 0 ~.
\erl{cc3}
We see that the commutation relations \rf{VR} define a subalgebra
$\g_R :=\sp{X_i^0}\iso \sp{X_i}$, because
$C_{ij}^\al$ must vanish to satisfy the condition \rf{cnd}, as was originally concluded in
\cite{iw}. Note that
\rf{cc3} tells us that $I^0 = \sp{X_\al^0}$ is an abelian subalgebra,
whereas \rf{CR} tells us that $I^0$ is an ideal of $\g^0$.

The contractions which are not of the \iw type are called {\em generalized \iw contractions}.
In the present paper we shall encounter one example of \iw contractions and two examples of
generalized \iw contractions.

To give the reader an intuitive understanding of the above definitions and notation, let
us consider the famous example of contracting the Lorentz algebra to the Galilean algebra:
Let $e_{ij}$ denote a basis of $4\x 4$  matrices, defined by
$(e_{ij})_{kl}=\delta_{ik}\delta_{jl}$. They have the following commutation relations
\be
[e_{ij} , e_{st}] = \delta_{js}e_{it} -\delta_{it}e_{sj}~, \quad i,j,s,t=1, 2, 3, 4.
\el{ee}
We define the three contracting boosts by
\be
B^\eps_i :=  \eps^2 \, e_{i4} +  e_{4i} =  \eps \,\left(\eps \, e_{i4} +
\frac 1 \eps \, e_{4i} \right)  = :  \eps \, f_\eps (B_i) ~, \quad i=1, 2, 3.
\el{boost}
These commute as follows
\be
[B^\eps_i , B^\eps_j] = \eps^2 \, [ e_{i4} , e_{4j}] = \eps^2\, ( e_{ij}-e_{ji})
=: - \eps^2\, L_{ij} \Ra 0~.
\el{bb}
which shows how the the Lorentz algebra $\so(3,1)$  for $\eps =1$ is contracted to the
Galilei algebra, which is the Euclidean algebra $\e(3)$, in which the limits of the
boosts $B_i^0=e_{4i}$ generate an abelian ideal.

\section{The hydrogen algebra $\hyt$ of $H_0$ }
\label{review}

Instead of the six generators $\Lb$ and $\Ab$ for the 3-D Kepler problem,
only three generators are conserved for  the 2-D
Kepler problem \cite{lf}. These are the third component of
angular momentum $L_3$ and two components of
the Runge-Lenz vector $\Ab$:
\br
L  & \equiv & L_3 := xp_y-yp_x= p_\ph \aand \cr
\Ab &:=& (A_1,A_2)=  L p_y\ \xbbh-L p_x\ \ybbh - m\al \ \rbbh
\erl{g0}
In the following I shall use the following notation:
\be
\fbox{\large~$h_0 \equiv -2mH_0~, \qquad
h \equiv -2mH \aand
\veps \equiv -2mE~.$~}
\el{not}
For simplicity and also in order to compare my results with those
of \cite{lf}, I shall use from now on  {\em Poisson brackets} instead
of commutation relations.
But I shall nevertheless refer sometimes to these Poisson brackets as commutators.

The Poisson brackets of the above generators are
\br
\{L, A_1\} &=& A_{2}~, \cr
\{A_2,L\} &=& A_1~, \where h_0:=-2mH_0     \lb{cr0}\\
\{A_1,A_2\} &=& h_0 L~,                    \nn
\er
The loop algebra \hyt is spanned by the following generators
\be
L^{(2n)} := h_0^n L \aand
A_i^{(2n+1)} := h_0^n A_i \quad (i=1,2) \aand n\ge 0~.
\el{ln0}
I call the upper index the {\em grade} of the corresponding operator.
According to the above construction,
every multiplication by $h_0$ raises the grade of the
generators by 2.
With the commutators \rf{cr0} the set
\be
\hyt := \{A_1^{(2n+1)}, A_2^{(2n+1)}, L^{(2n)} ~|~ n \ge 0\}
\el{lc0}
becomes a closed Lie algebra, which is a subalgebra of the affine
Kac-Moody algebra $A_1^{(1)}$.

\sbs{The factor algebra $\hyt/I(E)$ }

The three standard finite-dimensional algebras, $\so(3),
\so(2,1)$ and $\e(2)$, can be recovered from \hyt, as in \cite{dsd,dd}, as follows:
First we define an energy-dependent ideal of \HHt  by
\be
I(E) := (H_0-E)~ \HHt = (h_0-\veps) \HHt, \where \veps:=-2mE
\el{ideal0}
Next, we define the {\em energy-dependent} factor algebra $\HHt /I(E)$ relative to the above
ideal. This factor algebra
consists of three elements or classes,
\be
\HHt /I(E) =\{\Ac_1^\veps,\, \Ac_2^\veps,\, {\cal L}^\veps\}~,
\el{fa0}
which obey the following commutation
relations
\be
\fbox{\large $\{{\cal {L}}^\veps, \Ac_1^\veps \} = \Ac_2^\veps,
\quad \{\Ac_2^\veps,{\cal {L}}^\veps\} = \Ac_1^\veps \aand
\{\Ac_1^\veps, \Ac_2^\veps\}=\veps~{\cal {L}}^\veps~.$}
\el{basicE}
The commutation relations \rf{basicE} are exactly those of \rf{cr0}, except that the
operator $h_0$ in \rf{cr0} is now replaced by the numerical parameter $\veps$.
This is what physicists usually obtain by {\em directly} replacing the Hamiltonian $H$
by its energy eigenvalue $E$.

The above classes can be identified by their representatives, as follows
\be
\Ac_1^\veps = A_1+ I(E)~, \quad \Ac_2^\veps = A_2+ I(E)~, \aand
{\cal L}^\veps = L+ I(E)~.
\el{class0}
To see why each {\em  `basic element'} becomes a representative of its class, we recall
that quite generally an ideal $I$ of an algebra $\g$ acts additively as the zero
element of the factor algebra $\g/I$. In our case, this fact yields
the following equivalence relation in $\HHt /I(E)$,
\be
h_0^n X_i \equiv \veps^n X_i \quad \mod(I(E))~, \quad
\el{mod0}
where $X_i$ is a basic generator, \ie the element which generates the whole
infinite 'tower' $\{h_0^n X_i ~|~ n=0,1,\ldots \}$~.
The above equivalence relation can be proved easily, as follows
\be
h_0^n X_i - \veps^n X_i = (h_0^n -\veps^n) X_i
= (h_0-\veps) \left(\sum_{k=0}^{n-1}
\veps^{n-1-k} h_0^k \right)  X_i \in I(E)~.
\el{proof}
Equation \rf{mod0} tells us that in the factor algebra we can replace every element
$h_0^n X_i \in \HHt$ by $\veps^n X_i$, which is simply a numerical multiple of $X_i$.
Hence, in $\HHt/I(E)$ we can replace every element in the tower
$\{h_0^n X_i ~|~ n=0,1,\ldots \}$ by a single element $X_i$, so that $\HHt/I(E)$
is a finite-dimensional algebra generated by the $X_i$, which
in our case are the three elements given in \rf{class0}.\\
Note that {\em the Hamiltonian} $H_0$ {\em by itself is NOT an element of the ideal} $I(E)$.

\sbs{Contraction of the factor algebra $\hyt/I(E)$}

It  is easy to check that the map
\br
f_\veps(\sqrt{\sgn(\veps)}L_i) &=& \frac 1 {\sqrt{|\veps|}}~ \Ac_i^\veps ~, \quad i=1,2~,\cr
f_\veps(L_3)&=& {\cal {L}}^\veps ~,
\er
defines an isomorphism between the algebras $\so(3), \so(2,1)$ and the factor algebra
$\hyt/I(E)$ for $ \veps <0,\ \veps>0$. Hence, by treating $\veps = -2m E$ as a contraction
parameter $\eps$, the classes
$ \Ac_1^\veps,  \Ac_2^\veps $ and  ${\cal {L}}^\veps $ with the commutation relations
\rf{basicE} can be regarded as the generators of a {\em contracting} algebra
$\g^\eps$ (see Sec. 2), for $\veps \ne 0$ (!).

For $\veps \raz$ the algebras $\hyt/I(\veps)$ are contracted to $\hyt/I(0)$, whose commutation
relations follow from \rf{basicE}
\be
\{  {\cal L}^0, \Ac_1^0 \} = \Ac_2^0   ~,\quad
\{ \Ac_2^0, {\cal L}^0 \} =  \Ac_1^0 ~, \quad
\{ \Ac_1^0, \Ac_2^0\}= 0~,
\el{e2}
Since these are the commutation relations of  the Euclidean algebra $\e(2)$,
it follows that $\hyt/I(0)\iso \e(2)$.
Since $f_\eps$ is an isomorphism for $\veps\ne 0$, we conclude that a contraction of
$\hyt/I(\veps)$ for the non-broken Hamiltonian $H_0$ in \rf{h3} is the same as the well-known
contraction of $\so(3)$ and $\so(2,1)$ to the
Euclidean algebra $\e(2)$, as $\veps \raz$.
Note that the number of generators remains the same after contraction. In the
present case, the contraction is of the \iw type.

In the next two sections we shall see that the Factor algebras associated with the `broken
Hamiltonian' $H$ of \rf{h4} yield two  contractions of the generalized \iw type.

\section{The loop algebra $\Lf_1(\beta)$ of $H$ in \rf{h4}}

For the Hamiltonian \rf{h4} there exist a generalized conserved
Runge-Lenz vector \cite{lf}, which is given by
\br
\Mb & \equiv & \Mb(\beta) :=\Ab - m\beta \sqrt{r} \sin (\ph/2) ~\hat{\phib}(\ph) \cr
&=& \left(\frac {p_\ph^2} r -m\beta \right)~ \rbbh(\ph) -
\left(p_r p_\ph~+~ m\beta \sqrt{r} \sin (\ph/2) \right) ~\hat{\phib}(\ph)~.
\erl{M}
Note that $\Mb(0)= \Ab$~.
The commutator of the two components
of $\Mb$ in \rf{M} yield a third conserved quantity, which I shall denote by
$S$ (It is called $-I$ in \cite{lf}) :
It is defined by \cite{lf}
\be
S := \{M_1, M_2\}= h ~p_\ph - m\beta (p_r r^{1/2} \sin (\ph/2) +
p_\ph r^{-1/2} \cos (\ph/2) )~,
\el{S}
The commutators of $S$ with $M_i$ are \cite{lf}
\be
\{S, M_1\} =h~ M_2 \aand N_1:= \{M_2,S\} = h M_1 -m^2 \beta^2/2 ~.
\el{smm}
We can summarize the above commutators, as follows
\be
\{ S, N_1\} = h^2 M_2~, \quad \{M_2,S\} = N_1 \aand
\{N_1, M_2\} = h S ~.
\el{basic}
Therefore, I call the following  three generators,
{\em `basic generators'}
\be
N_1,\quad M_2, \aand S,
\el{basicg}
because they can yield a closed algebra by multiplying them with powers
of $h$ as in \rf{ln} below.
{\em The above basic generators were chosen, such that none of them vanishes
nor blows up for $H=0$}~.

As before, since $H$ commutes with the basic generators,
we can close the algebra in \rf{basic} by including the following generators
\be
 h^n M_2~,  \quad
 h^n N_1 ~,\quad
h^n S~, \quad n\ge 0~.
\el{ln}
It is interesting to note that by commuting the basis generators $N_1, M_2$ and $S$,
among themselves and with their commutators, we can never produce $hM_2$. This means
that it is possible to obtain a closed algebra even without $h M_2$.
Nevertheless, I included $h M_2$ in \rf{ln} in order to obtain a closed algebra
which is generated by the basic generators over the polynomial ring $R[h]$.

A crucial step in identifying the algebra generated by the operators in \rf{ln} is to
{\em assign grades} to each operator, because for
Lie algebras of the Kac-Moody type
the sum of the grades (which I am writing as upper indices) must be
conserved under commutation. It is easy to check, that the following
identification of the grades is consistent
\be
\fbox{\large $
M_2^{(2n+1)} := h^n M_2~,  \quad
N_1^{(2n+3)} := h^n N_1~,\quad
S^{(2n+2)} := h^n S~, \quad~ n\ge 0~.$}
\el{ln2}
For example, using \rf{basic} we obtain
\[
\{N_1^{(2m+3)}, M_2^{(2n+1)}\} = h^{m+n} \{N_1,M_2\} = h^{m+n+1} S =S^{(2m+2n+4)}~.
\]
Therefore the above infinite generators span the following graded Loop algebra
of the Kac-Moody type,
\be
\Lf_1(\beta) := \{M_2^{(2n+1)}, N_1^{(2n+3)}, S^{(2n+2)} ~|~ n \ge 0\}~.
\el{lc}
Note that the basic generators are graded, as follows
\[
M_2 =M_2^{(1)}~,  \qquad   S= S^{(2)} \aand  N_1= N_1^{(3)}~.
\]

\sbs{The factor algebra $\Lf_1(\beta)/I_1(E,\beta)$}

As before, the factor algebra
\be
\Lf_1(\beta) /I_1(E,\beta) =\{\Mc_2^\veps,\, \Nc_1^\veps,\, \Sc^\veps\},
\el{fa1}
relative to the following energy-dependent ideal
\be
I_1(E,\beta) := (H-E)~ \Lf_1(\beta) = (h-\veps) \Lf_1 (\beta)~,
\where  \veps=-2mE~.
\el{ideal}
has three classes, which commute as follows
\be
\fbox{\large $\{\Sc^\veps, \Nc_1^\veps \}
= \veps^2 \Mc_2^\veps, \quad \{\Mc_2^\veps,\Sc^\veps\}
= \Nc_1^\veps \aand \{\Nc_1^\veps, \Mc_2^\veps\}=\veps~\Sc^\veps~.$}
\el{basicE1}

\sbs{Contraction of the factor algebra $\Lf_1/I_1(E,\beta)$}

In the present case we need a different map
\br
f_\veps(\sqrt{\sgn(\veps)}L_1) &=& \Nc_1^\veps/|\veps|^{3/2}=:
\widehat{\Nc_1}^\veps \cr
f_\veps(\sqrt{\sgn(\veps)}L_2) &=& \Mc_2^\veps/|\veps|^{1/2}=:
\widehat{\Mc_2}^\veps \lb{norm} \\
f_\veps(L_3)&=&  \Sc^\veps/|\veps|=: \widehat{\Sc}^\veps~, \nn
\er
which again defines an isomorphism between the algebras $\so(3), \so(2,1)$ and the factor
algebra $\Lf_1/I_1(E,\beta)$ for $ \veps <0,\ \veps>0$.
The three generators $\widehat{\Nc_1}^\veps,
\widehat{\Mc_2}^\veps $ and $\widehat{\Sc}^\veps$ may be called `normalized' generators.

The contraction of the factor algebras
$\Lf_1/I_1(E,\beta)$ yields $\Lf_1/I_1(0,\beta)$, whose commutation relations follow
from \rf{basicE1}. They are given by
\br
[\Nc_1^0, \Mc_2^0] &=& \lim_{\veps\raz}  \veps^{3/2+1/2-1} \Sc^\veps= 0  \cr
~[\Sc^0, \Nc_1^0] &=& \lim_{\veps\raz}  \veps^{1+3/2-1/2} \Mc_2^\veps = 0  \lb{so3w} \\
~[\Mc_2^0, \Sc^0 ] &=& \lim_{\veps\raz}  \veps^{1/2+1-3/2} \Nc_1^\veps = \Nc_1^0 ~.
\erl{ccl}
These are the commutation relations of the Heisenberg-Weyl algebra $\h_3=\w_1$,
as we can see by using the following map
\[
\Mc_2^0\ra \partial_x, \qquad \Sc^0\ra x, \aand \Nc_1^0\ra 1~.
\]

It is important to note that in the factor algebra $\Lf_1(\beta)/I_1(E,\beta)$ we are NOT
allowed to replace the $h$ in $N_1$ of  \rf{smm} by $\veps$, since neither $h$ nor
$(h-\veps) M_1$ are elements of the ideal $I_1(E,\beta)$. Hence, $N_1$ is independent of
$E$ and thus it should  NOT be replaced by
the constant $-m^2 \beta^2/2$ for $E=0$.

\section{A second loop algebra $\Lf_2(\beta)$ of $H$ in \rf{h4} }

In this section I show that a different choice of the basic generators yields
different contractions. Instead of the three generators in \rf{basicg} I now
choose the basic generators, as follows
\be
N_1, \qquad N_2 := h M_2 \aand S~.
\el{newb}
The choice of $N_2$ in \rf{newb} may seem unjustified.
But I chose it nevertheless in order to illustrate
how we can
obtain different contractions by simply removing some generators from
the {\em same} loop algebra.

The choice \rf{newb} would seem less strange, had I kept
the phase angle $\ga$ in \rf{h4} arbitrary : In this case I would
have obtained
\br
\tilde N_1 &:=& \{\tilde M_2, \tilde S\} = \tilde h \tilde M_1
-\half  m^2 \beta^2 \cos \ga    \aand \crn
\tilde N_2 &:=& \{\tilde S,\tilde M_1\} =\tilde h \tilde M_2
- \half m^2 \beta^2\sin \ga
\er
where the tilde over the
quantities denote the quantities of the previous section, but
with $\ga\ne 0$. Hence, for $\ga$ arbitrary, the $\tilde N_1, \tilde N_2$
and $\tilde S$ would have seemed to be the natural
choice for the basic generators.
In fact, these generators were originally chosen by Leach and Flessas
\cite[Eq. (3.4.5)]{lf}
as the symmetry generators of the Hamiltonian \rf{h4} for $E\ne 0$.
However, for $E=0$
they made a different choice, and chose the following linear
combinations of $\tilde N_1$ and $ \tilde N_2$
\br
N_1 &=& \cos \ga \tilde N_1+ \sin \ga \tilde N_2=  h M_1 -\half  m^2 \beta^2
\cr
M_2 &=& \frac 1 h ( \sin \ga \tilde N_1- \cos \ga \tilde N_2) ~.
\erl{sec}
We see that their second choice \rf{sec} corresponds exactly to the
generators which I used in Sec. 4,  by setting $\ga=0$
from the beginning.
This explains why they were able to obtain the algebra $\h_3=\w_1$ as the symmetry
algebra for $E=0$; for $E\ne 0$ it does not matter which linear combinations
one chooses: one always obtain $\so(3)$ or $\so(2,1)$.

The generators in \rf{newb} commute, as follows
\be
\{N_1, N_2\} =h^2 S, \quad \{N_2,S\} = h N_1, \aand \{S, N_1\} =h~ N_2
\el{cr2}
Following the same procedure as before, the following operators
\br
N_1^{(2n+3)} &:=& h^n N_1~,  \cr
N_2^{(2n+3)} &:=& h^n N_2~,  \for n\ge 0~,\lb{ln3} \\
S^{(2n+2)} &:=& h^n S~, \nn
\er
yield the following Loop algebra, provided one uses
the grading in \rf{ln3}
\be
\Lf_2 := \{N_1^{(2n+3)}, N_2^{(2n+3)}, S^{(2n+2)} ~|~ n \ge 0\}
\el{lc2}

\sbs{The factor algebra $\Lf_2(\beta)/I_2(E,\beta)$}

The factor algebra in this case consists also of three classes, namely
\be
\Lf_2(\beta) /I_2(E, \beta) =\{\Nc_1^\veps, \Nc_2^\veps, \Sc^\veps\}~,
\el{fa2}
where
\be
I_2(E, \beta) := (H-E) \Lf_2(\beta) = (h-\veps) \Lf_2(\beta)~.
\el{ideal2}
These classes commute as follows
\be
\fbox{\large$\{\Nc_1^\veps, \Nc_2^\veps\}=\veps^2~\Sc^\veps~,
\quad \{\Nc_2^\veps, \Sc^\veps\} = \veps \Nc_1^\veps \aand
\{\Sc^\veps, \Nc_1^\veps \} = \veps \Nc_2^\veps~.$}
\el{basicE2}
Hence, in this case we obtain for $\veps\raz$ a contraction
of $\so(3)$ and $\so(2,1)$ to an abelian algebra, which I denote by $R^3$. This is a
generalized \iw contraction.

Note that if $N_2=hM_2$, as defined in \rf{newb}, then
$N_2^{(2n+3)}=M_2^{(2n+3)}$,
so that $\Lf_2$ is just a subalgebra of
$\Lf_1$, with just the element $M_2$ removed, \ie
\be
\Lf_2=\Lf_1 \backslash M_2^{(1)} =\Lf_1 \backslash  M_2~.
\el{lt}
Again note that in the factor algebra $\Lf_2(\beta)/I_2(E,\beta)$ we are NOT
allowed to replace $N_2=hM_2$ by $\veps M_2$, since
$(h-\veps) M_2$ is NOT an element of the ideal  $I_2(E,\beta)$, because \rf{lt} tells
us that $M_2 \not \in \Lf_2$. Thus, the class $\Nc_2^0 = N_2 + I_2(0) \ne I_2(0)$,
which means that the contracted factor algebra  $\Lf_2(\beta)/I_2(0,\beta)$ remains
three-dimensional, as it should. Note that with the formal factor-algebra construction
every one of the three generators is well defined and will not vanish in the limit
$\veps \raz$,
so that this realization is {\em saved} \cite{sal}.
{\em  In contrast, if instead we follow the standard procedure
and work directly with the generators $N_1, N_2$ and $S$ and just replace the $h$
by $\veps$, then $N_2=h M_2$ will become $N_2=\veps M_2$ and thus it will vanish in the
limit $\veps\raz$}, so that $N_2$ will not be saved.

\section{Summary and conclusions}

In the present paper I constructed two Kac-Moody loop algebras $\Lf_1(\beta)$ and
$\Lf_2(\beta)$. The second algebra $\Lf_2(\beta)$ was studied simply to show that
one has the freedom of constructing more than one loop algebra from the conserved
constants of motion, $M_1, M_2, S$ and $H$. These two infinite-dimensional algebras
are operator-valued and thus do NOT depend on energy $E$.

To study contractions I first constructed $E$-dependent factor algebras,
in order to obtain finite-dimensional algebras out of the infinite-dimensional ones.
As I explained in Eq. \rf{mod0}, this construction enables us to replace all the higher
generations $X_i^n:=h^n X_i$ by $\veps^n X_i$, so that within the factor algebras
all the generators $X_i^n$ become
numerical multiples of the basic generators $X_i=X_i^0$. In particular, for $E=0$ we
obtain $\veps^n X_i =0$ for $n\ge 1$.

To avoid any misunderstanding, I want to emphasize again that I am NOT contracting the
infinite-dimensional Kac-Moody loop algebras, $\H_2, \Lf_1(\beta)$ and $\Lf_2(\beta)$;
{\bf I am only contracting their (3-dimensional) factor algebras},
$\H_2/I(E), \Lf_1(\beta)/I_1(E,\beta)$ and $\Lf_2(\beta)/I_2(E,\beta)$, by using the
energy $E$ as the contraction parameter.
It is interesting that although all the three factor
algebras  are isomorphic to $\so(3)$ and $\so(2,1)$ for $E<0$ and
$E>0$, they contract for $E\ra 0$ to three different algebras
$\e(2), \h_3=\w_1$ and $R^3$, which are also 3-dimensional.
The first contraction is of the \iw type while the other two are of the generalized \iw
type. In all these contractions  the dimension of the algebras is preserved,
since the factor algebras do not change their dimensions as $E\raz$.
These contractions are summarized in table 1.\\
\begin{table}[ht]
\label{tab:1}
\vspace{.4cm}
\begin{center}
\begin{tabular}{|c|| c   || c | c| c|}
\hline 
 & & & & \\
Hamiltonian & Factor algebra &  $E < 0 $ &  $E = 0 $
& $E < 0 $    \\ & & & & \\
\hline 
& & & & \\
$H_0$ in \rf{h3} &
$ \hyt /I(E)$ & $\so(3)$ & $ \e(2)$   & $\so(2,1)$ \\& & & & \\
$H$ in \rf{h4} & $\Lf_1(\beta)/I_1(E,\beta)$ &  $\so(3)$ & $ \h_3=\w_{\!1}$   & $\so(2,1)$  \\
 & & & & \\
$H$ in \rf{h4} & $\Lf_2(\beta)/I_2(E,\beta)$ &  $\so(3)$ & $ R^3 $   & $\so(2,1)$  \\
& & & & \\
\hline 
\end{tabular}
\caption{The three factor algebras $ \protect\hyt /I(E)$
and $\Lf_i/I_i$
of the loop algebras \protect\hyt
 and $\Lf_i$ relative to the corresponding
energy-dependent ideals $I(E)$ and $I_i(E,\beta)$. For $E\ne 0$ all three
factor
algebras are isomorphic to $\so(3)$ for $E<0$ and to $\so(2,1)$ for $E>0$,
but yield {\em different contractions}
for $E \ra 0$~.}
\end{center}
\end{table}
The effect of symmetry breaking $H(\beta)$ manifests itself differently in
in the standard and the Kac-Moody treatments: In the standard procedure, which was
followed by Leach \etal \cite{gl,lf}, the symmetry algebras for $H_0$ and $H(\beta)$
are exactly the
same, namely $\so(3)$ and $\so(2,1)$.
The effect of symmetry breaking manifests itself only for $E\ne 0$.

In contrast, as we shall now see, the Loop algebras $ \Lf_1 $ and $\Lf_2$ for the
`broken Hamiltonian' $H(\beta)$
are {\em smaller} than the hydrogen algebra $\H_2$ for $H_0$ (irrespective of the energy!).
They are smaller by two and three elements,
respectively, thereby revealing the symmetry breaking:

To compare  $ \Lf_1 $ and $\Lf_2$ with $\H_2$, I define two maps, as follows:
$f_1: \Lf_1 \mapsto \hyt$, defined by
\br
f_1(N_1^{(2n+3)}(\beta)) &:=& A_1^{(2n+3)}~, \cr
f_1(M_2^{(2n+1)}(\beta))   &:=&A_2^{(2n+1)} ~,\for n\ge 0 ~, \lb{f1map} \\
f_1(S^{(2n+2)}(\beta))  &:=& L^{(2n+2)}~,  \nn
\er
and $f_2: \Lf_2 \mapsto \hyt$, defined by
\br
f_2(N_1^{(2n+3)}(\beta)) &:=& A_1^{(2n+3)} ~,\cr
f_2(N_2^{(2n+3)}(\beta))  &:=& A_2^{(2n+3)} ~, \for n\ge 0 ~, \lb{f2map} \\
f_2(S^{(2n+2)}(\beta))  &:=&  L^{(2n+2)}~, \nn
\er
It is easy to check that these two maps, which keep the grades of the generators unchanged,
define isomorphisms from $\Lf_1$
and $\Lf_2$ onto subalgebras of \hyt of codimension 2 and 3, respectively.
Hence,
\be
\HHt = \left\{\ba{ll} f_1(\Lf_1(\beta)) \cup \{L, A_1\}~, &\aand \\[3mm]
f_2(\Lf_2(\beta)) \cup \{L, A_1,  A_2\}~.& \ea \right.
\el{cup}
Thus, we can conclude that symmetry breaking of the type \rf{h4}
reduces the loop algebra \hyt of the original system $H_0$ by only finite number
of generators.
By constructing the corresponding factor algebras, I obtained different
contractions depending on the missing terms (see Table 1).

By defining the $\veps$-dependent ideals and constructing
the factor algebras, we are essentially replacing each infinite-dimensional
{\em `tower'} $ \{ X_i^n \}$ by one element $X_i^{n_i^{min}}$ which has the
{\em  lowest grade}. {\em By removing generators
from the original loop algebra, we increase the grade of the
corresponding basic generators.}
This in turn increases the powers of the contraction
parameter $\veps$ which multiply the structure constants of the original
algebra $\g$, which is being contracted.

The results obtained in the present paper suggest a {\em general
procedure for defining contractions via Kac-Moody formalism}, as follows:

\begin{itemize}

\item {Start with of a finite dimensional Lie algebra $\g$, which may be
graded, via $s$-dimensional automorphism, as follows
\be
\g = \bigoplus_{k=0}^{s-1}~ \g_{k} ~, \with [\g_{i}, \, \g_{j}]\subseteq \g_{i+j}~,
\el{grading}
where the indices are modulo $s$~.}
\item {Then consider the {\em positive} subalgebra of
a general (twisted or untwisted) loop algebra of a finite dimensional algebra $\g$,
\be
\Lf =\left\{ \bigoplus_{k=0}^{s-1}~ z^{sn+k} \otimes \g_{k} ~ \mid~ n \ge 0 ~\right\} ~,
\el{lcg}
where $t$ may be a scalar or an operator which commutes with all the generators of
$X_i \in \g$.}

\item Then remove some generators from $\Lf$, and make sure that the {\em remaining set}
$\Lf_R$ yields a subalgebra of $\Lf$. This is not automatic: see, for example,
the conditions in \rf{nnn} below. Then make sure that the set
\be
I_R(\veps) = (z-\veps) \Lf_R
\el{lcr}
is an ideal of $\Lf_R$, since for some choices $(z-\veps) \Lf_R$ is not a subalgebra of
$\Lf_R$~.

\item Finally, define the factor algebras $\Lf_R/I_R(\veps)$, which will be
isomorphic to one or two
real forms of $\g$, depending on the sign of parameter $\veps$. The   $\veps$ can be
used as a contraction parameter. One may get different contractions
for the same original algebra $\g$ as $\veps \raz$, depending on the
removed generators.\\

For example, we can define subalgebras of $\H_2$ by
\be
\H_2(n_1, n_2, n_3):= \sp{h^{n_1+n} A_1, h^{n_2+n} A_2,
h^{n_3+n} L_3, n\ge 0}
\el{hcon}
if the $n_i$ satisfy the following conditions
\be
n_1+n_2-n_3+1\ge 0~, \quad n_3+n_1-n_2 \ge 0, \quad n_3+n_2-n_1\ge 0~.
\el{nnn}
In particular, as I showed explicitly in \rf{f1map} and \rf{f2map}, the loop algebras
$\Lf_i$  are isomorphic
to the following subalgebras of $\H_2$, and thus give us intuitive physical
realizations of the formal definition in \rf{hcon}:
\be
\Lf_1 \iso \H_2(1, 0, 1) \aand \Lf_2 \iso \H_2(1, 1, 1)
\ee
In these subalgebras of $\H_2$ the conditions \rf{nnn} are  clearly satisfied.\\

The conditions \rf{nnn} follow from two different arguments:
\begin{enumerate}

\item The generators of the subalgebra $\H_2(n_1, n_2, n_3)$ commute, as follows
\br
[h^{n_1}A_1,h^{n_2}A_2] &=& h^{n_1+n_2+1}L_3, \mbox{\quad hence~}  n_3 \le n_1+n_2+1 \cr
[h^{n_3}L_3,h^{n_1}A_1]&=& h^{n_3+n_1}A_2,  \mbox{~~\quad hence~}  n_2 \le n_3+n_1 \cr
[h^{n_3}L_3,h^{n_2}A_2]&=& -h^{n_3+n_2}A_1,  \mbox{\quad hence~}  n_1 \le n_3+n_2
\erl{cc7}
The conditions \rf{nnn} are necessary to ensure that the  r.h.s. of the above commutators
are elements of $\H_2(n_1, n_2, n_3)$.

\item
In the factor algebra $\H_2(n_1, n_2, n_3)/(~(h-\veps)\H_2(n_1, n_2, n_3)~)$
only the generators with lowest grade are linearly independent. Their commutators are
\br
[\eps^{n_1}A_1,\eps^{n_2}A_2] &=& \eps^{n_1+n_2-n_3+1}  (\eps^{n_3} L_3) \cr
[\eps^{n_3}L_3,\eps^{n_1}A_1]&=& \eps^{n_3+n_1-n_2} (\eps^{n_2}A_2) \cr
[\eps^{n_3}L_3,\eps^{n_2}A_2]&=& -\eps^{n_3+n_2-n_1} (\eps^{n_1} A_1)
\erl{cc8}
Hence, in order for the r.h.s. of the above three equations to exist as $\eps\raz$,
the exponents of $\eps$ must be non-negative. This requirement yields
exactly the same conditions on the $n_i$ as
those given in \rf{cc7}, which were necessary for the existence of subalgebras of $\H_2$.

\end{enumerate}

\item More generally, given an $N$-dimensional semisimple algebra $\g$, we can define
subalgebras $\g_{\bn}$ by
\be
\g_{\bn} := \sp{h^{n_i} X_i~|~n\ge 0 \aand i=1,2,\ldots N}~.
\el{gbn}
Instead of an operators $h$, with $[h, X_i]=0$, we can also use a formal variable $z$.

These subalgebras yield well-defined contractions via the factor-algebra
$\g_{\bn}/(\,(h-\eps) \g_{\bn}\,)$,
provided the $n_i$ satisfy the general condition \rf{cnd},  namely
$\eps^{n_i+n_j-n_k} C_{ij}^k < \infty$~.

\end{itemize}

\begin{figure}[htbp]
\centering
\epsfig{file=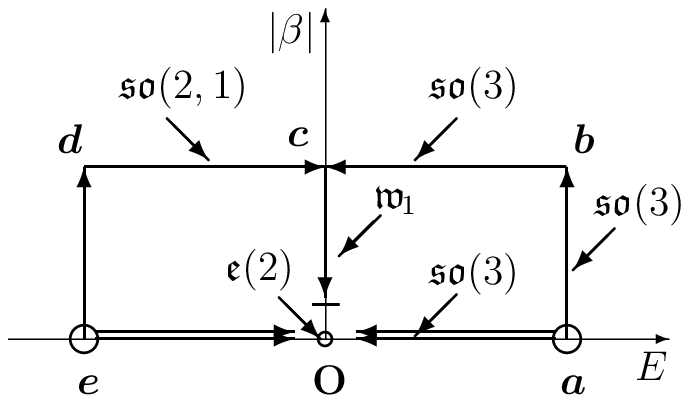, width=8.5cm} 
\caption{The figure illustrates the `{\em DC hysteresis}' in the $(E, \beta)$
parameter plane, by comparing the contraction limits $E\raz$ of the factor
algebras $\hyt/I(E)$ and $\Lf_1/I_1(E,\beta)$:
If we contract the factor algebras $\hyt/I(E)$ of $H_0$ along the horizontal
energy $E$-axis, which corresponds to $\beta=0$, we obtain $\e(2)$.
This contraction is indicated by the double arrows
($ \eb \Ra {\bf O} \Leftarrow \ab $).
In contrast, for $\beta\ne 0$ the contraction of $\Lf_1/I_1(E,\beta)$ of $H$ yields
the Weyl algebra $\Lf_1/I_1(0,\beta)=\h_3= \w_{\! 1}$, as illustrated by $\db \ra \cb \leftarrow
\bb$.
Finally, taking the limit of $\Lf_1/I_1(0,\beta)$ as $\beta\raz$ downwards along the
vertical $|\beta|$-axis to the origin $(E,\beta)= (0,0)$ leaves the algebra $\h_3=\w_{\! 1}$
unchanged. Thus, the two
paths originating in $\ab$ yields different limits:
$\so(3)\simeq \hyt/I(E) \simeq \Lf_1(\beta)/I_1(E,\beta)\ra \Lf_1(\beta)/I_1(0,\beta)
\simeq \Lf_1(0)/I_1(0,0)\simeq \h_3=\w_{\! 1}  \ne \e(2)\simeq \hyt/I(0) \Leftarrow \hyt/I(E) $.}
\label{fig.1}\end{figure}

Finally, we note that if we {\em first} take the limit $\beta \raz$ in the
{\em `deformed Hamiltonian'} $H(\beta)$ of \rf{h4} we recover the original
Hamiltonian $H_0$ and thus obtain the symmetry algebra $\H_2$ and consequently the contraction to $\e_2$. In contrast, 
if we construct the loop symmetry algebras 
$\Lf_i(\beta) $ for $\beta \ne 0 $ first, then the $\Lf_i(\beta) $ (and also their factor algebras) remain unchanged as we 
let  $\beta \raz$, and thus we do NOT go back
to \hyt (and its factor algebras). I call this phenomenon the {\bf DC
(deformation-contraction) hysteresis}, since {\em we obtain different contractions
depending on the \underline{order} of taking the limits $E\raz $ and $\beta\raz$.}
The subtlety of the DC hysteresis, which yields $\h_3=\w_1$ instead of $\e_2$
is illustrated in Fig. 1.

\section*{Acknowledgments}

I am happy to thank my daughter, Claudia Daboul, for reading the manuscript
and making useful comments.
It is also a pleasure to thank the members of CCF, and in particular
Bernardo Wolf, for their hospitality.


\edd

\newpage


\def\prosnfigure{\begin{picture}(10,5.5) 
\unitlength 7mm

\thicklines

\put(9.5,0.4){$E$}

\put(5.7,2.8){$\w_{\! 1}$}
\put(5.5,2.8){\vector(-1,-1){.45}}
\put(6.5,1.8){$\so(3)$}
\put(6.9,1.6){\vector(-1,-1){.45}}
\put(6.5,4.3){$\so(3)$}
\put(6.9,4.1){\vector(-1,-1){.45}}
\put(3.8 ,1.8){$\e(2)$}
\put(4.5,1.55){\vector(1,-1){.45}}

\put(4.15,5.2){$|\beta|$}
\put(8.6,3.7){{\bb}}
\put(8.4 , 0.1){{\ab}}
\put(9.5 , 2.8){$\so(3)$}
\put(9.2, 2.8){\vector(-1,-1){.45}}
\put(4.8,.0){{\bf O}}      \put(5, 1){\circle{.7}}

\put(4.6,3.8){{\cb}}
\put(1.4 , 0.1){{\eb}}
\put(-.8 , 2.8){$\so(2,1)$}
\put(.9, 2.8){\vector(1,-1){.45}}

\put(1.5,3.5){\vector(1,0){3.5}}   
\put(1.65,1){\vector(1,0){2.9}}   
\put(1.65,1.1){\vector(1,0){2.9}}   
\put(1.5,1.){\vector(0,1){2.5}}   
\put(1.5,1){\circle{.4}}
\put(1.1,3.7){{\db}}

\put(8.5,3.5){\vector(-1,0){3.5}}
\put(8.35,1){\vector(-1,0){2.9}}\put(8.35,1.1){\vector(-1,0){2.9}}
\put(8.5,1.){\vector(0,1){2.5}}
\put(8.5,1){\circle{.4}}
\put(5,3.5){\vector(0,-1){1.9}} 
\put(4.8,1.5){\line(1,0){.4}} 

\thinlines
\put(5,1){\vector(0,1){4.4}}
\put(0,1){\vector(1,0){10}}


\end{picture}}


\begin{figure}[ht]
\hspace{-30mm} \psfig{file=prosnfig.ps}
          \label{fig.1}
\end{figure}

\edd
\begin{thebibliography}{99}

\bibitem{pauli} W. Pauli, Z. Physik. 36, 336 (1926); English transl.:
in {\em Sources of Quantum Mechanics},
B. L. van der Waerden (Ed.)(North-Holland, Amsterdam, 1967),

\bibitem{sud}
L. I. Schiff, Quantum Mechanics, 3rd edition (McGraw-Hill, 1968);
A. Sudbery, {\em Quantum Mechanics and the Particles of Nature} (Cambridge
University Press, 1986);  H. Goldstein, {\em Classical Mechanics}
(Addison-Wesley, 1980).

\bibitem{dsd} J.\ Daboul, P.\ Slodowy and C. \ Daboul,
Phys. Lett.  B 317,  321 (1993);
C. \ Daboul, J.\ Daboul, and P.\ Slodowy, {\it The Dynamical Algebra
of the Hydrogen Atom as a Twisted Loop Algebra}, Proceedings of the XX
International Colloquium on " Group Theoretical Methods in Physics",
Osaka, July 4-9, 1994,
A.\ Arima, T.\ Eguchi and N. \ Nakanishi (Eds.), (World Scientific,
Singapore, 1995) p. 175-178 (hep-th/9408080).
These papers contain a short review of the basic concepts of the
Kac-Moody algebras.

\bibitem{dd} C.\ Daboul and J. \ Daboul, Phys. Lett. B 425, 135 (1998);
J.\ Daboul, C. \ Daboul and P.\ Slodowy, {\it Affine Kac-Moody
Algebras and the D-dimensional Hydrogen Atom}, in ``Symmetry and
Structural Properties of Condensed Matter", eds. T. Lulek, W. Florek and
B. Lulek (World Scientific, Singapore, 1997)  p. 338-347.

\bibitem{kac} R. V. Kac, {\em Infinite Dimensional Lie Algebras}, 3rd ed.,
Cambridge University Press, Cambridge, 1990.

\bibitem{fs}  J. Fuchs and C. Schweigert, {\em  Symmetries, Lie Algebras
and Representations} (Cambridge University Press, 1997) 

\bibitem{win} P. Winternitz, J. A. Smorodinsky, M. Uhlir and I. Fris,
Sov. J. Nucl. Phys. 4, 444 , 1967; For a recent discussion of the
relevant potentials, see
M. B. Sheftel, P. Tempesta and P. Winternitz, J. Math. Phys. 42, 659
(2001);
P. Tempesta, A. V. Turbiner and P. Winternitz, J. Math. Phys. 42, 4248 (2001)

\bibitem{sen} T. Sen, J. Math. Phys. 28, 2841 (1987).

\bibitem{gl} V. M. Gorringe and P. G. L. Leach,
J. Austral. Math. Soc. Ser. B 34, 511-522 (1993).

\bibitem{lf} P. G. L. Leach and G. P. Flessas, J. Nonlinear Math. Phys.
10, 340-423 (2003).

\bibitem{kmp} E. G. Kalnins, W. Miller Jr and G. S. Pogosyan, J. Phys. A, 33,
4105 (2000) and 33, 6791 (2000); E. G. Kalnins, J. M. Kress, W. Miller Jr
and G. S. Pogosyan, J. Phys. A 34, 4705 (2001).

\bibitem{iw}  E. In\"on\"u and E. P. Wigner, Proc. Nat. Acad. Sci. USA 39,
510 (1953)

\bibitem{gil} R. Gilmore, {\em Lie Groups, Lie Algebras and Some of Their
Applications}, Wiley, 1974, and references therein.

\bibitem{cla} C. Daboul, {\em Deformationen und Degenerationen von
Liealgebren und Liegruppen}, Doctoral dissertation,
University of Hamburg, 1999.~ (in German)

\bibitem{sal} E. J. Saletan, J. Math. Phys. 2, 1 , 1961.


\end{thebibliography}
